\documentclass[twocolumn,twoside,slac_two]{revtex4}
\usepackage{graphicx}
\usepackage{fancyhdr}
\begin{document}

\title{Optical Link ASICs for the LHC Upgrade}

%

\author{K.K. Gan, H.P. Kagan, R.D. Kass, J.R. Moore, D.S. Smith}
\affiliation{Department of Physics, The Ohio State University, Columbus, OH 43210, USA}

\begin{abstract}
We have designed three ASICs for possible applications in the optical links of a new layer of pixel detector in the ATLAS experiment for the first phase of the LHC luminosity upgrade. The ASICs include a high-speed driver for the VCSEL, a receiver/decoder to decode the signal received at the PIN diode to extract the data and clock, and a clock multiplier to produce a higher frequency clock to serialize the data for transmission. These ASICs were designed using a 130 nm CMOS process to enhance the radiation-hardness. We have characterized the fabricated ASICs and the submission has been mostly successful. We irradiated the ASICs with 24 GeV/c protons at CERN to a dosage of 70 Mrad. We observed no significant degradation except the driver circuit in the VCSEL driver fabricated using the thick oxide process in order to provide sufficient voltage to drive a VCSEL. The degradation is due to a large threshold shifts in the PMOS transistors used.
\end{abstract}

\maketitle

\thispagestyle{fancy}


\section{Introduction}
The Large Hadron Collider (LHC) at CERN, Geneva, Switzerland, will be upgraded in two phases, resulting in ten times higher luminosity. The ATLAS detector is one of the two major experiments at the LHC. For the first phase of the luminosity upgrade, the plan is to install a new layer of pixel detector inside the present pixel detector to compensate for the expected degradation due to the intense radiation. We have designed three ASICs for possible applications in the optical links for the new pixel layer.

The prototype ASICs were fabricated using a 130 nm CMOS 8RF process.  The ASICs contain three circuit blocks~\cite{Arms}; a VCSEL driver (optimized for operation at 640 Mb/s and 3.2 Gb/s), a PIN receiver with a clock and data recovery circuit capable of operation at 40, 160, or 320 Mb/s, and a clock multiplier (two versions) designed to operate at 640 Mb/s. The clock multiplier is needed to produce a higher frequency clock to serialize the data for transmission. All circuitries were designed following test results and guidelines from CERN on radiation tolerant design in the 130 nm process used~\cite{Faccio}. We have characterized the ASICs in the lab followed by an irradiation with 24 GeV protons at CERN. The results are summarized below. All ASICs tested were packaged for the irradiation and therefore we experienced some speed degradation due to the added stray capacitance of the packaging.

\section{VCSEL Driver Chip}
The two VCSEL driver circuits in the test chip have similar architecture; one optimized for 640 Mb/s with higher drive current and the other for 3.2 Gb/s with lower current.  Each consists of a 1.5 V LVDS receiver circuit, a 1.5 to 2.5 V logic converter circuit, and a 2.5 V VCSEL driver circuit.  Both VCSEL drivers allow for adjustable bias and modulation currents and contain circuitry to reduce switching noise on the power supply lines.  The LVDS receiver was designed using the standard thin oxide transistors that operate with a 1.5 V supply. The use of the of the thin oxide transistors allowed us to not only achieve higher bandwidth over the thick oxide transistors but also produce a circuit that could be used by other members of our community developing chips which operate with a single 1.5 V supply.  We used a 2.5 V supply for the driver circuitry because most commercially available VCSELs require bias voltages greater than 2 V to produce suitable output optical power.  The driver portion of the chips were therefore designed using the thick oxide transistors available in the 130 nm process.   Previous results from CERN have shown that thin oxide transistors designed using conventional layout techniques exhibit radiation tolerance suitable for SLHC applications. However, conventionally designed thick oxide transistors are not suitably radiation tolerant.  All of the thick oxide transistors were therefore designed using an enclosed structure.

\begin{figure}[h]
\centering
\includegraphics[width=80mm]{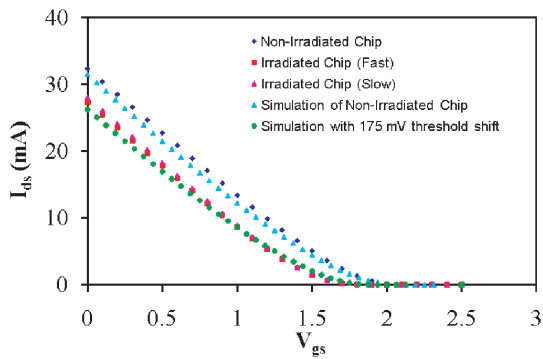}
\caption{The IV curves of a PMOS transistor in the VCSEL driver before and after irradiation. Also show is the simulation before and after irradiation with 175 mV of threshold shift.} \label{IV}
\end{figure}

Four prototype ASICs were packaged and the performance was satisfactory up to 1 Gb/s. The performance at higher speeds could not be sufficiently evaluated due to the packaging parasitics.  These tests also verified the operation of the LVDS receiver up to 1 Gb/s.  During the irradiation, each driver ASIC was connected to a 25~$\Omega$ resistor instead of a VCSEL to allow testing of the degradation of the chip alone. The duty cycle of the output signal and the current consumption of the LVDS receiver remained constant during the irradiation. However, we observed significant decrease in the VCSEL driver circuit current consumption and the output drive current. It should be noted that the decrease in the drive current could be compensated by adjusting the control current but this option was not used during the irradiation in order to study the degradation under the same condition. Post-irradiation analysis indicates that there is no other significant degradation and the decreases in the VCSEL driver current is now understood. Figure~\ref{IV} shows the IV curves of a PMOS transistor fabricated in the thick oxide technology before and after irradiation. The IV characteristic before the irradiation is well reproduced by the simulation. A shift is observed after the irradiation and is well reproduced by the simulation after including a 175 mV threshold shift. The threshold shift is significantly smaller in the NMOS transistors. The present current mirror in the driver circuit uses both type of transistors and thus is sensitive to the different threshold shifts. The plan is to use PMOS transistors only in the future design.

\section{PIN Receiver and Decoder}

The PIN receiver/decoder contains a trans-impedance amplifier, limiting amplifier, bi-phase mark (BPM) decoder with clock recovery, and LVDS drivers to send the decoded data and recovered clock off the ASIC.  All transistors are fabricated with thin oxide, allowing the ASIC to run with a 1.5 V power supply.  The clock and data recovery is accomplished using a delay locked loop that contains networks of switchable capacitors in the delay stages allow for the three operating frequencies, 40, 160, and 320 Mb/s. However, due to insufficient time to optimize the design, the ASICs operate at somewhat lower speed and require higher threshold currents to achieve a low bit error rate (BER). Other than these two limitations, the performance of the ASICs is satisfactory, including clock jitter, duty cycle, rise/fall time, and high/low levels of the LVDS drivers.

\begin{figure}[h]
\centering
\includegraphics[width=80mm]{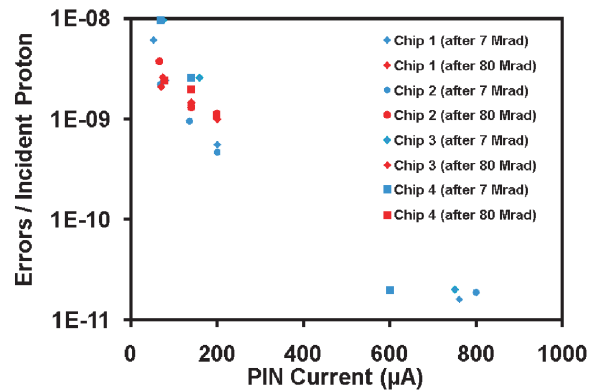}
\caption{BER as a function of the PIN current threshold for a PIN receiver/decoder coupled a PIN diode at two doses, 7 and 80 Mrad.} \label{BER}
\end{figure}

For the irradiation, the ASICs were tested in two different setups of four ASICs each. The first setup was purely electrical while the second setup involved a PIN diode so that we could decouple the electrical and optical degradations. In both setups, the decoded data were transmitted to the control room using 20 m of coax. In the first setup, 40 Mb/s BPM signals were transmitted over 20 m of coax to the ASICs. The long cable precluded testing at higher speed. In the second setup, we sent 40 Mb/s BPM signal via a fiber to a PIN diode coupled to an ASIC. For both setups, we observed single event upset during the spill but observed no degradation in the threshold for $\sim$1 error/s as a function of dosage. We also monitored the BER vs. PIN current during the irradiation. The BER decreased with larger PIN current and was higher for an ASIC coupled to a PIN diode as expected (Fig.~\ref{BER}). The current consumption was constant during the irradiation. Detailed post-irradiation analysis reveals no significant degradation in the overall performance.

\section{Clock Multiplier}

The clock multiplier circuits consist of charge pump/ring oscillator phase locked loops (PLL) with dividers in the feedback loop.  One circuit performs a frequency multiplication of 16 and the other a multiplication by 4, both yielding a 640 MHz clock. The multipliers are designed using thin oxide transistors and are powered by a single 1.5 V supply.  The multiplier circuits share a common 50~$\Omega$ driver and LVDS receiver enabling testing of only one multiplier at a time.  An additional switching network allows the recovered clock from the PIN receiver prototype to be routed to either multiplier to allow for testing of how the received signal jitter is coupled to the multiplied clock.

Four multipliers were packaged for the irradiation. All ASICs functioned well with clock jitter of $<$ 8 ps (0.5\%). During the irradiation, we observed that the clocks of two ASICs lost lock and power cycling was needed to resume operation at 640 MHz. This problem is not yet understood. Detailed post-irradiation analysis also reveals no significant degradation in the overall performance.

\section{Summary}

We have designed prototype ASICs using the 130 nm process to enhance the radiation-hardness. The submission has been mostly successful. We irradiated the ASICs to a dose of 70 Mrad and observed no significant degradation except in the VCSEL driver. Post-irradiation analysis indicates that there is a significant threshold shift in the PMOS transistors fabricated in the thick oxide technology for the operation at 2.5 V to drive the VCSEL. An improved version of the ASIC will be submitted implementing what we learned from the study.

\begin{acknowledgments}
This work was supported in part by the U.S. Department of Energy under contract No. DE-FG-02-91ER-40690. The authors are indebted to M. Glaser for his tireless assistance in the use of the T7 irradiation facility at CERN.
\end{acknowledgments}


\end{document}